\documentclass{webofc}
\usepackage[varg]{txfonts}   
\begin{document}
\title{Exotic polyquark states and their properties in QCD}
%
%


\author{\firstname{Dmitri} \lastname{Melikhov}\inst{1,2,3}\fnsep\thanks{\email{dmitri_melikhov@gmx.de}}}

\institute{Institute for High Energy Physics, Austrian Academy of Sciences, Nikolsdorfergasse 18, A-1050 Vienna, Austria 
\and
D.~V.~Skobeltsyn Institute of Nuclear Physics, M.~V.~Lomonosov Moscow State University, 119991, Moscow, Russia
\and
Faculty of Physics, University of Vienna, Boltzmanngasse 5, A-1090 Vienna, Austria}

\abstract{%
We formulate rigorous criteria for selecting diagrams to be taken into account in the analysis of possible 
polyquark (tetra or penta) poles in QCD. The central point of these criteria is the requirement that the Feynman diagrams for the relevant Green functions 
contain four-quark (or five-quark in the pentaquark case) intermediate states and the corresponding cuts. It is shown that some diagrams which ``visually'' seem to contain 
the four-quark cuts, turn out to be free of these singularities and therefore should not be taken into account when calculating the 
tetraquark properties. We then consider large-$N_c$ QCD which in many cases provides qualitatively correct picture of hadron properties 
and discuss in detail the tetraquark states. For the ``direct'' and the ``recombination'' four-point Green functions, which may potentially 
contain the tetraquark poles, we formulate large-$N_c$ consistency conditions which strongly restrict the behaviour of the 
tetraquark-to-ordinary meson transition amplitudes. In the end, these conditions allow us to obtain constraints on width of the possible  
tetraquark states at large $N_c$. We report that both flavor-exotic and cryptoexotic (i.e., flavor-nonexotic) tetraquarks, 
if the corresponding poles exist, have a
width of order $O(1/N_c^2)$, i.e. they should be parametrically narrower compared to the ordinary $\bar qq$ mesons with a width of
order $O(1/N_c)$. Moreover, for flavor-exotic states, the large-$N_c$ consistency conditions require two narrow
flavor-exotic states, each of these states coupling dominantly to one specific meson-meson channel.
}
\maketitle

\section{Motivation}
In the recent years, many candidates for exotic tetraquark mesons (most of the states with charmonium or 
bottomonium like structure, i.e. with the hidden charm or the hidden beauty) and a few candidates for pentaquark states have been reported (see \cite{olsen,ali}).
To be more precise, the experiments have observed the resonance-like structures in the region near the two-hadron thresholds tempted  
to be interpreted as the new hadron poles. The precise status of these structures is still not fully understood 
(whether they correspond to the poles or are just complicated kinematical near-threshold effects), and the interest in the 
theoretical understanding of these states is incredible. The two main questions concerning the exotic states to be answered are the following: 
(i) do the scattering amplitudes have new poles (in the complex plane), and 
(ii) if the new poles exist, what is the structure of the corresponding hadrons, i.e. whether they are dominantly molecular-like 
states---bound states in the Van-der-Waals potential between colorless hadrons, 
or they are bound states of two or more colored objects in a confined potential. 
Answering these questions starting from the QCD Lagrangian is an immensely difficult problem. 

To provide the theoretical understanding of exotic states, one may consider QCD with a large number of colors $N_c$ [i.e., ${\rm SU}(N_c)$ gauge
theory for large $N_c$] with a simultaneously decreasing coupling $\alpha_s\sim1/N_c$ \cite{largeNc1,largeNc2}: in the sector of the ordinary hadrons, 
large-$N_c$ considerations are known to provide qualitative explanations of many phenomena of real strong interactions corresponding to $N_c=3$. 
A specific feature of the large-$N_c$ QCD, which simplifies the analysis of the QCD Green functions, is that at the leading order, 
large-$N_c$ QCD Green functions receive contributions only from planar diagrams (with an infinite number of gluon exchanges) and as the result 
have only non-interacting mesons as intermediate hadron states; if tetraquark bound states emerge at all, they may emerge only in subleading large-$N_c$ 
diagrams \cite{coleman}. For many years, this fact was believed to be an argument providing the theoretical explanation of
the non-existence of exotic tetraquarks. However, it was noticed in \cite{weinberg}, that even if the exotic tetraquark bound states appear
in the subleading diagrams, the crucial question is the width of these objects: if narrow, they might still be well observed in nature.
The conclusion in \cite{weinberg} obtained on the basis of the large-$N_c$ counting, that if tetraquark states exist, they may be as narrow 
as the ordinary mesons; the latter are known to have width $\sim1/N_c$. The question of the expected width of the exotic states 
(if such poles indeed exist) has been further addressed in \cite{knecht}, discussing the dependence of the width of tetraquark mesons 
on their flavor structure. Finally, \cite{maiani} reported for the cryptoexotic tetraquarks an even
smaller width of order $N_c^{-3}$. 

A detailed analysis of the exotic states in large-$N_c$ QCD was performed by our group in \cite{lms2017}. The main content of the present talk 
is based on these results. 
 
\section{Diagram selection criteria}
In order to draw a conclusion about the width of a possible tetraquark pole, one has to understand which QCD diagrams 
(in the general case, and not only at large-$N_c$) may lead to the appearance of this pole and thus should be included in the analysis. 

The crucial property of such diagrams is in fact obvious: since we are going to study bound state containing four quarks, the appropriate 
QCD diagrams should contain four-quark intermediate states and the corresponding cuts in the variable $s$ if we expect to observe a 
tetraquark pole in $s$ \cite{cohen}. 
An unambiguous way to establish the presence (or the absence) of such a four-particle $s$-cut is to apply the Landau equations \cite{landau} 
which allow one to find the location of all singularities of the Feynman diagrams. A detailed application of the Landau equations 
shows that some of the diagrams which visually seem to contain the four-quark intermediate state and which thus have been attributed to the 
tetraquark pole in some of the previous analyses, in fact do not contain the necessary four-particle cut and should be therefore excluded 
from the analysis of the tetraquark properties. 
This observation has crucial consequences for the properties of the possible tetraquark poles at large-$N_c$:
According to our findings \cite{lms2017}, the tetraquark width at large $N_c$ does not depend on
its flavor structure: both flavor-exotic and flavor-nonexotic tetraquarks have parametrically the same width of order $1/N_c^2$. 

Let us study the four-point Green functions of colorless bilinear quark currents (i.e. interpolating currents for ordinary mesons) 
of various flavor content. 

The general four-point function depends on six kinematical variables: the four momenta squared of the external currents, 
$p_1^2$, $p_2^2$, $p_1^{\prime2}$, $p_2^{\prime2}$, $p=p_1+p_2=p_1'+p_2'$, and the two Mandelstam variables $s=p^2$ and $t=(p_1-p_1')^2$. 
A bound state in the $s$-channel corresponds to a pole in the variable $s$ in the four-point function.  

Now, the diagrams which
potentially may contribute to the tetraquark pole at $s=M_T^2$, should satisfy the following two criteria:
\begin{enumerate}
\item 
The diagram should have a nontrivial (i.e., non-polynomial) dependence on variable $s$.
\item
The diagram should have four-quark intermediate states in the $s$-channel and the corresponding $s$-cuts starting at
$s=(m_1+m_2+m_3+m_4)^2$, where $m_i$ are the masses of the quarks which are suspected to form the tetraquark bound state. 
The presence or absence of this cut should be established by solving the Landau equations for the corresponding diagram.
\end{enumerate}
Obviously, not all diagrams in the perturbative expansion of Green functions satisfy the above criteria. It is therefore convenient 
to decompose these diagrams into two sets: diagrams of the first set either do not depend on $s$ or have no
four-quark cut in the $s$-channel and thus cannot be related to the possible tetraquark poles; 
diagrams of the second set satisfy both above criteria~and thus contribute to the potential tetraquark pole.

Having at hand the above criteria, we study separately two cases: 
tetraquarks of an exotic flavor content (i.e., built up of quarks of four different flavors, $\bar q_1q_2\bar q_3q_4$) and
cryptoexotic tetraquarks (i.e. those of the flavor content $\bar q_1q_2\bar q_2q_3$,~having the same flavor as ordinary mesons have). 
The necessity to treat flavor-exotic and cryptoexotic cases separately arises~from the different topologies of the appropriate QCD diagrams in 
these two cases.

\section{Flavour-exotic tetraquarks}
Let us consider a bilinear quark current $j_{ij}=\bar q_i\hat O q_j$ producing a meson $M_{ij}$ of flavor content $\bar q_iq_j$
from the vacuum, $\langle 0|j_{ij}|M_{ij}\rangle=f_{M_{ij}}$. Here, $\hat O$ is a combination of Dirac matrices corresponding to
the meson's spin and parity. We omit all Lorentz structures as they are not important for our analysis. 
Important for us is that at large $N_c$, the meson decay constants $f_{M}$ scale~as $f_{M}\sim\sqrt{N_c}$.

In the case of four-point functions of bilinear currents involving
quarks of four different flavors, denoted by $1,2,3,4$, one encounters two types of Green functions: the ``direct'' functions
$\Gamma^{\rm(dir)}_{{\rm I}}=\langle
j^\dagger_{12}j^\dagger_{34}j_{12}j_{34}\rangle$ and
$\Gamma^{\rm(dir)}_{{\rm II}}=\langle
j^\dagger_{14}j^\dagger_{32}j_{14}j_{32}\rangle$, and the
``recombination'' functions $\Gamma^{\rm(rec)}=\langle
j^\dagger_{12}j^\dagger_{34}j_{14}j_{32}\rangle$ and
$\Gamma^{\rm(rec)\dag}$.
It turns out that the analysis of the four-quark singularities of the ``direct'' Green functions is rather straightforward, 
whereas the analysis of the ``recombination'' Green functions has subtleties. 

\subsection{Direct channel}
\begin{figure}[h!]
\centering
\includegraphics[width=10.5cm]{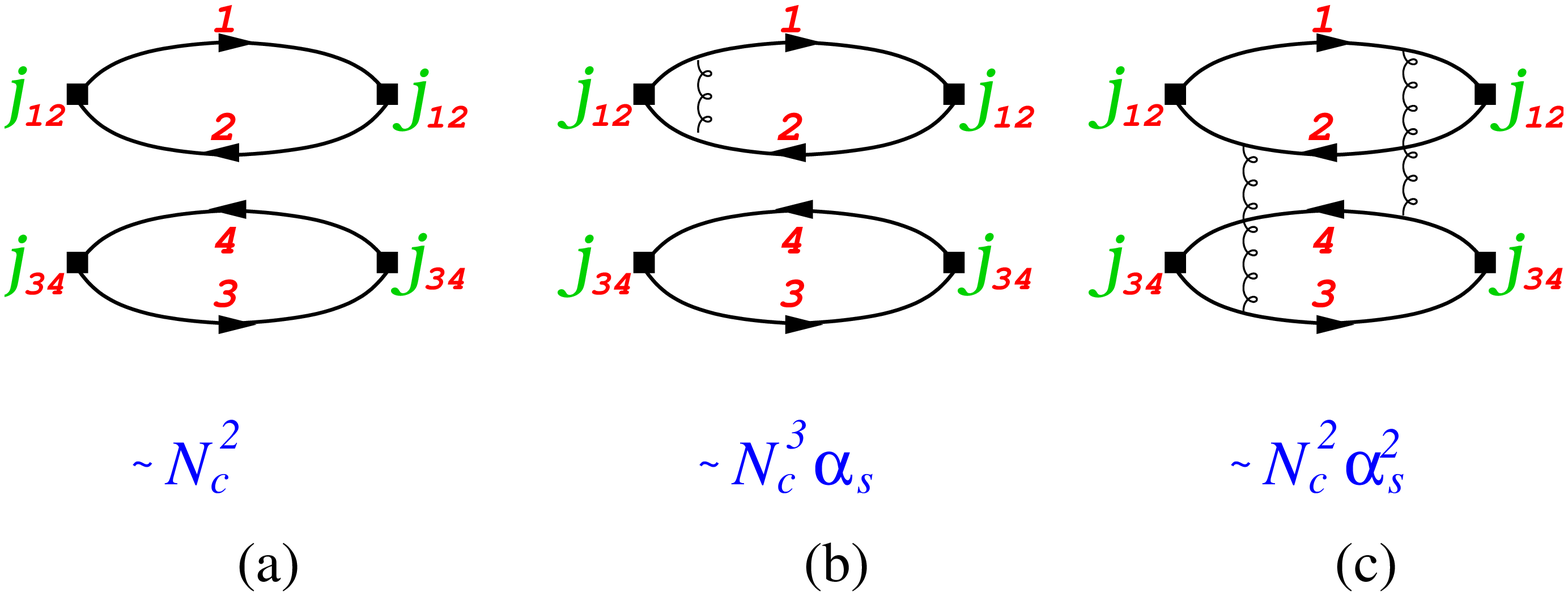}
\caption{\label{Fig:dir}
Perturbative expansion of the ``direct'' Green functions $\Gamma^{\rm(dir)}_{I}$.}
\end{figure}
Figure~\ref{Fig:dir} shows the perturbative expansion of the
direct correlator $\Gamma^{\rm(dir)}_{{\rm I}}$. Similar diagrams
defined by evident flavor rearrangements describe the correlator
$\Gamma^{\rm(dir)}_{{\rm II}}$. Obviously, not all these diagrams
satisfy our above criteria for diagrams that potentially contain a
tetraquark pole. For instance, the diagrams in
Fig.~\ref{Fig:dir}(a,b) do not depend on $s$. The leading
large-$N_c$ diagram which depends on $s$ and also has a four-quark
$s$-cut is given by Fig.~\ref{Fig:dir}(c). The diagrams of this
type are therefore the leading large-$N_c$ diagrams of interest to us.

\subsection{Recombination channel}
\begin{figure}[!h]
\centering
\includegraphics[width=11cm]{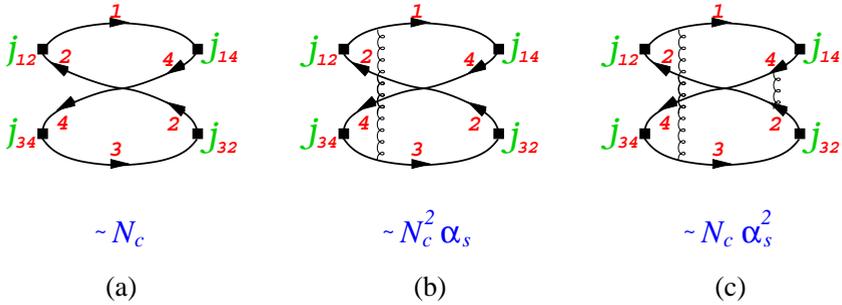}
\caption{\label{Fig:rec1}Diagrams for the recombination Green function $\Gamma^{(\rm rec)}$.}
\end{figure}

The analysis of the recombination channel is more involved: among the diagrams in Fig.~\ref{Fig:rec1}, the first two diagrams
(a,b) do depend on $s$; however, in spite of their appearance,
they have no four-quark $s$-cut (and in fact no four-quark cuts at all).

\begin{figure}[!h]
\centering
\includegraphics[height=10cm]{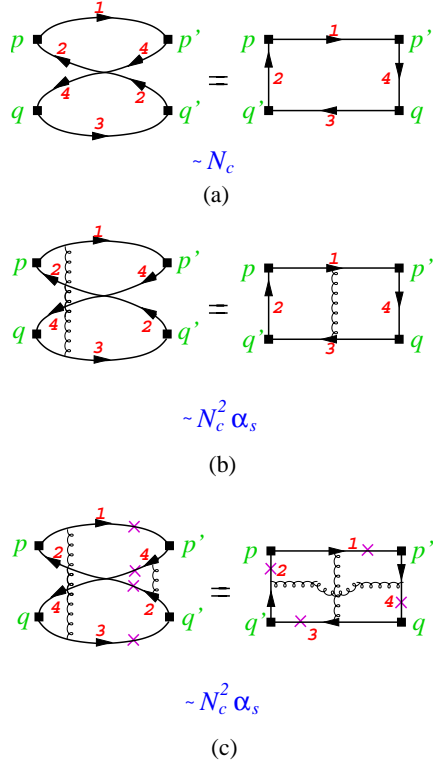}
\caption{\label{Fig:rec2}
Diagrams for the recombination Green function $\Gamma^{(\rm rec)}$ displayed as box diagrams.}
\end{figure}

The easiest way to see this is to redraw these diagrams as the usual box diagram with gluon exchanges, see Fig.~\ref{Fig:rec2}: 
The $s$-cut of the diagrams on the l.h.s. corresponds to the $u$-cut of the box diagrams on the r.h.s.
For the r.h.s. diagrams (a) and (b), the absence of four-quark singularities is obvious and of course can be confirmed by solving the Landau equations. 
(Moreover, the r.h.s. diagrams have no two-quark $u$-cuts). Therefore, the l.h.s. diagrams (a) and (b) have no four-quark singularities at all. 
Let us emphasize that planar gluon exchanges in the diagrams on the r.h.s. of (a) and (b) do not change this property. 
In order to develop a four-quark cut, one needs a non-planar gluon exchange: a non-planar diagram in the r.h.s. of Fig.~\ref{Fig:rec2} (c) develops a four-quark $u$-cut 
when the crossed propagators go on-shell. Respectively, this corresponds to the four-particle $s$-cut of the l.h.s. diagram in Fig.~\ref{Fig:rec2} (c). 
Finally, the leading large-$N_c$ diagram exhibiting the four-quark $s$-cut is the nonplanar diagram in Fig.~\ref{Fig:rec2}(c). Its four-quark cut with 
the threshold at $s=(m_1+m_2+m_3+m_4)^2$ may be verified directly by solving the Landau
equations.

Based on the established leading large-$N_c$ diagrams possessing four-particle intermediate states, 
we find the following leading large-$N_c$ behaviour of the Green functions $\Gamma_{T}$ possibly containing a pole
corresponding to some tetraquark $T$:
\begin{eqnarray}
\label{T1} 
\Gamma^{\rm(dir)}_{{\rm I},T}&=&\langle
j^\dagger_{12}j^\dagger_{34}j_{12}j_{34}\rangle=O(N_c^0),
\nonumber\\
\Gamma^{\rm(dir)}_{{\rm II},T}&=&\langle
j^\dagger_{14}j^\dagger_{32}j_{14}j_{32}\rangle=O(N_c^0),
\\
\Gamma^{\rm(rec)}_T&=&\langle
j^\dagger_{12}j^\dagger_{34}j_{14}j_{32}\rangle=O(N_c^{-1}).
\nonumber
\end{eqnarray}
These diagrams potentially contain the tetraquark pole, although
the actual existence of this pole is still a conjecture. Now, let
us assume that narrow resonances (i.e., resonances with widths
vanishing for large $N_c$) show up at the~lowest possible $1/N_c$
order and that the resonance mass $M_T$ remains finite at large
$N_c$. The fact that the ``direct'' and the ``recombination'' Green functions of flavour-exotic currents have
different large-$N_c$ behaviours, leads to the conclusion that a
single pole cannot saturate the leading behaviour of both the direct and the recombination Green functions;  
we need~at least two exotic poles, denoted by $T_A$ and $T_B$: one state, $T_A$, couples stronger to the
$M_{12}M_{34}$ channel, while another state, $T_B$ couples stronger to the $M_{14}M_{32}$ channel.

Truncating the poles corresponding to the external mesons and retaining explicitly only the tetraquark poles, 
we~get
\begin{eqnarray}
\label{T4b}
\Gamma_{{\rm I},T}^{(\rm dir)}&=&O(N_c^0)=f_M^4\left(\frac{|A(M_{12}M_{34}\to T_A)|^2}{p^2-M^2_{T_A}}
+\frac{|A(M_{12}M_{34}\to T_B)|^2}{p^2-M^2_{T_B}}\right)+\cdots,
\nonumber\\
\Gamma_{{\rm II},T}^{(\rm dir)}&=&O(N_c^0)=f_M^4\left(\frac{|A(M_{14}M_{32}\to T_A)|^2}{p^2-M^2_{T_A}}+\frac{|A(M_{14}M_{32}\to T_B)|^2}{p^2-M^2_{T_B}}\right)+\cdots,
\\
\Gamma_{T}^{(\rm rec)}&=&O(N_c^{-1})=f_M^4\left(\frac{A(M_{12}M_{34}\to T_A)A(T_A\to M_{14}M_{32})}{p^2-M^2_{T_A}}\right.
\nonumber\\
&&\hspace{2cm}\left.+\frac{A(M_{12}M_{34}\to
T_B)A(T_B\to M_{14}M_{32})}{p^2-M^2_{T_B}}\right)+\cdots.\nonumber
\end{eqnarray}
At this point we have to take into account the scaling relation $f_M\sim\sqrt{N_c}$. Now, since we are interested 
in the properties of the possible tetraquark poles with {\bf finite mass} at large $N_c$, the equations (\ref{T4b}) have the following solution:
\begin{eqnarray}
\label{T4b2}
&& A(T_A\to M_{12}M_{34})=O(N_c^{-1}),\qquad A(T_A\to M_{14}M_{32})=O(N_c^{-2}),\nonumber\\
&& A(T_B\to M_{12}M_{34})=O(N_c^{-2}),\qquad A(T_B\to M_{14}M_{32})=O(N_c^{-1}).
\end{eqnarray}
The widths $\Gamma(T_{A,B})$ of the states $T_A$ and $T_B$ are
determined by the dominant channel. We therefore conclude that both states have parametrically the same widths,  
$\Gamma(T_{A,B})=O(N_c^{-2})$.

So far we have ignored the mixing between $T_A$ and $T_B$.
Introducing their mixing parameter $g_{AB}$, we get additional
contributions to the above Green functions. Most restrictive for
$g_{AB}$ is the recombination function, for which mixing provides
the additional contribution
\begin{eqnarray}
\label{mixing}
\Gamma_{T}^{(\rm rec)}=O(N_c^{-1})
=f_M^4\left(\frac{A(M_{12}M_{34}\to T_A)}{p^2-M^2_{T_A}}g_{AB}
\frac{A(T_B\to M_{14}M_{32})}{p^2-M^2_{T_B}}\right)+\cdots.
\end{eqnarray}
Equations (\ref{T4b2}) and (\ref{mixing}) restrict the behaviour of
the mixing parameter in the following way: 
\begin{eqnarray}
g_{{}_{AB}}\le O(N_c^{-1}). 
\end{eqnarray}
Thus, the two flavor-exotic tetraquarks of the same flavor content do not
mix at large $N_c$.

\section{Cryptoexotic tetraquarks}
Let us now discuss tetraquarks with nonexotic flavor content, i.e.,
having the same flavor as the ordinary mesons. All confirmed tetraquark candidates observed experimentally belong to this class of states. 
The analysis proceeds along the same line as for the exotic states. There is however an important new ingredient compared to flavour-exotic states: 
diagrams of new topologies compared to the flavor-exotic case emerge. 

\subsection{Direct channel}
\begin{figure}[!h]
\centering
\includegraphics[width=8cm]{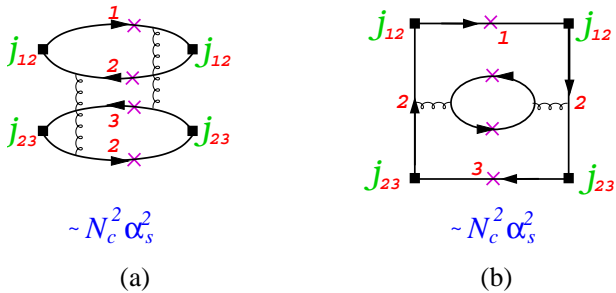}
\caption{\label{Fig:crypto_dir}
Diagrams contributing to
$\Gamma_{{\rm I},T}^{\rm(dir)}$. Crossed propagators denote on-shell particles contributing to the four-quark cut.}
\end{figure}

For the direct Green functions
$\Gamma^{\rm(dir)}_{({\rm I},{\rm II}),T}$ the new diagrams are shown in Fig.~\ref{Fig:crypto_dir}. Essentail for us is that 
the new diagrams do not change the
leading large-$N_c$ behaviour compared to the diagrams of the same topology as in the flavor-exotic case.

\subsection{Recombination channel}
\begin{figure}[!b]
\centering
\includegraphics[width=8cm]{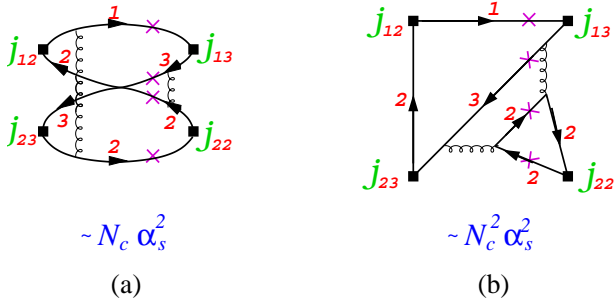}
\caption{\label{Fig:crypto_rec}
Diagrams contributing to the
recombination Green function $\Gamma_T^{\rm(rec)}$.}
\end{figure}

For the recombination functions, however, the situation changes
qualitatively: the new diagram, Fig.~\ref{Fig:crypto_rec}(b),
dominates at large $N_c$ and thus modifies the leading large-$N_c$
behaviour of $\Gamma^{\rm(rec)}_{T}$. We thus find
\begin{eqnarray}
\label{C1}
\Gamma^{\rm(dir)}_{{\rm I},T}=\langle
j^\dagger_{12}j^\dagger_{23}j_{12}j_{23}\rangle=O(N_c^0),\nonumber\\
\Gamma^{\rm(dir)}_{{\rm II},T}=\langle
j^\dagger_{13}j^\dagger_{22}j_{13}j_{22}\rangle=O(N_c^0),\\
\Gamma^{\rm(rec)}_T=\langle
j^\dagger_{12}j^\dagger_{23}j_{13}j_{22}\rangle=O(N_c^0).
\nonumber
\end{eqnarray}
In contrast to the flavor-exotic case, now both the direct and the
recombination Green functions have the same leading behaviour at
large $N_c$. As a consequence, one exotic state $T$ is already sufficient in order to
satisfy the expected large-$N_c$ behaviour of both Green functions.
The couplings of this state to the meson-meson channels are
\begin{eqnarray}
\label{C2}
&&A(T\to M_{12}M_{23})=O(N_c^{-1}),\nonumber\\
&&A(T\to M_{13}M_{22})=O(N_c^{-1}).
\end{eqnarray}
The exotic state $T$ is then expected to have coupling of the same order to both possible decay channel, i.e. no preferrable decay channel is seen.  
The width of this single cryptoexotic state is of order $\Gamma(T)=O(N_c^{-2})$.

If all its quantum numbers allow, $T$ can mix with the ordinary meson $M_{13}$. The restriction on the mixing parameter
$g_{TM_{13}}$ may be obtained, e.g., from the direct amplitude 
\begin{eqnarray}
\label{mixing2}
\Gamma_{{\rm I},T}^{(\rm dir)}=O(N_c^0)
=f_M^4\left(\frac{A(M_{12}M_{23}\to T)}{p^2-M^2_{T}}g_{TM_{13}}
\frac{A(M_{13}\to M_{12}M_{23})}{p^2-M^2_{M_{13}}}\right)+\cdots.
\end{eqnarray}
Taking into account that $A(M_{13}\to M_{12}M_{23})\sim
1/\sqrt{N_c}$ \cite{largeNc1,largeNc2}, we obtain for the mixing parameter 
\begin{eqnarray}
g_{TM_{13}}\le O(1/\sqrt{N_c}).
\end{eqnarray}
Thus, the cryptoexotic tetraquark and the ordinary meson with the same flavour do not mix at large $N_c$. 

\newpage
\section{Conclusions and outlook}
We have formulated a set of physically obvious and rigorous criteria for selecting Feynman diagram appropriate for the analysis of possible tetraquark 
(i.e. a bound state containing four quarks, independently of the structure of this bound state): namely, if one is going to study properties of the tetraquark, only
those diagrams, which contain four-quark intermediate states and the corresponding four-quark cuts in the appropriate variable should be taken into account. 
Diagrams which do not contain the four-quark intermediate states cannot contribute to the appearance of the tetraquark pole and should therefore be neglected. 

In particular, when one discusses four-point functions of the bilinear quark currents, describing the scattering of the ordinary $\bar qq$-mesons, 
one should consider only those contributions to the four-point Green functions which have a four-quark cut in the $s$-channel. 
The straightforward and rigorous way to establish these four-quark singularities is provided by the Landau equations. 
These simple criteria are quite general and are applicable in QCD for any $N_c$. 

Now, these general criteria for the diagram selection, combined with the large-$N_c$ limit, provide essential constraints on the properties of the possible 
exotic tetraquark mesons. Namely, using these criteria and requiring that
the {\bf narrow} poles contribute to the appropriate parts of the Green
functions at the leading large-$N_c$ order, we reported the following results: 
\begin{enumerate}
\item
One has important consistency conditions which arise from the large-$N_c$ behaviour of the direct and the recombination four-point Green functions. 
These consistency conditions restrict the number of exotic states necessary to saturate the leading large-$N_c$ behaviour of the four-point Green functions. 

\item 
For the flavor-exotic four-point Green functions (all four quarks of
different flavors $\bar q_1,\bar q_3,q_2, q_4$), their large-$N_c$ behaviour is not compatible
with the existence of only one flavor-exotic tetraquark bound state; the consistency condition requires two narrow
states $T_A$ and $T_B$ with~widths $\Gamma(T_A,T_B)=O(N_c^{-2})$.
Each of these tetraquarks dominantly couples to one meson-meson
channel; its coupling to the other meson-meson channel is
suppressed: $A(T_A\to M_{12}M_{34})=O(N_c^{-1})$, $A(T_A\to
M_{14}M_{32})=O(N_c^{-2})$, $A(T_B\to M_{14}M_{32})=O(N_c^{-1})$,
$A(T_B\to M_{12}M_{34})=O(N_c^{-2})$. The parameter describing the
mixing between~the two tetraquarks vanishes for large $N_c$ at
least like $1/N_c$.\item The large-$N_c$ behaviour of four-point
Green functions of nonexotic flavor content ($\bar q_1,\bar
q_2,q_2,q_3$) is compatible with the existence of a single narrow
cryptoexotic tetraquark, $T$, with width $\Gamma(T)=O(N_c^{-2})$.
This tetraquark couples parametrically equally to both two-meson
channels: $A(T\to M_{12}M_{23})=O(N_c^{-1})$, $A(T\to
M_{13}M_{22})=O(N_c^{-1})$. If quantum numbers allow, the
cryptoexotic tetraquark $T=\bar q_1q_3\bar q_2 q_2$ mixes with the
ordinary meson $M_{13}=\bar q_1q_3$. The corresponding mixing
parameter vanishes like $1/\sqrt{N_c}$, i.e., slower than that of
the flavor-exotic tetraquarks.
\item
The constraints on the width of the possible narrow exotic states at large $N_c$ are general and universal, i.e. they do not depend on the 
structure of the tetraquark pole. In particular, the constraints should be satisifed by both molecular and confined tetraquarks. 

\item
We discussed the consequences of the large-$N_c$ behaviour of four-point Green functions for narrow exotic states with the mass staying 
finite at large $N_c$. 
However, the derived constraints on the Green functions are quite general and cannot be violated by any exotic tetraquarks, including those 
with the masses that increases parametrically with $N_c$. In this case, however, one should take into account that the number of the open channels with many 
ordinary mesons in the final states also increases with $N_c$.  
Therefore, obtaining constraints on the width of the exotic states with the mass rising with $N_c$ needs a detailed treatment. 
\end{enumerate}
We would like to mention that, in principle, there is a possibility that narrow
tetraquarks do exist but appear only in $N_c$-subleading diagrams
with four-quark intermediate states, while they do not contribute
to the $N_c$-leading~topologies. This possibility seems rather
unnatural to us: if such pole exists at all, there should be some
special reason, not evident to us, why it does not appear in the
set of the appropriate leading large-$N_c$ diagrams. Nevertheless, also
in the $N_c$-subleading topologies one observes a difference in
the large-$N_c$ behaviour of the direct and the recombination
diagrams for the flavor-exotic case: for any $N_c$-subleading
topology, the direct diagrams are $N_c$-even, whereas the
recombination~diagrams are $N_c$-odd. Accordingly, if the latter
scenario is realized in nature, one still needs two flavor-exotic poles, although with parametrically smaller 
widths at large $N_c$.

\vspace{.5cm}\noindent{\it{\bf Acknowledgements.}}
I have pleasure to thank W.~Lucha and H.~Sazdjian for a fruitful collaboration on the subject of this talk.  
Valuable discussions with V.~Anisovich, T.~Cohen, M.~Knecht, B.~Moussallam,
O.~Nachtmann, and B.~Stech are gratefully acknowledged. The work was supported by the Austrian Science Fund (FWF) 
under project~P29028.

\end{document}